**Spectrometric method to detect exoplanets as another test to verify the invariance of light velocity**

B. R. Mushailov, V. S. Teplitskaya Sternberg Astronomical Institute, Moscow State University

Universitetsky pr., 13, Moscow 119992, Russia

**Abstract**

Hypothetical influences of variability of light velocity due to the parameters of the source of radiation, for the results of spectral measurements of stars to search for exoplanets are considered. Accounting accelerations of stars relative to the barycenter of the star - a planet (the planets) was carried out. The dependence of the velocity of light from the barycentric radial velocity and barycentric radial acceleration component of the star should lead to a substantial increase (up to degree of magnitude) semi-major axes of orbits detected candidate to extrasolar planets. Consequently, the correct comparison of the results of spectral method with results of other well-known modern methods of detecting extrasolar planets can regard the results obtained in this paper as a reliable test for testing the invariance of the velocity of light.

**Keywords**: Exoplanet system, spectrometric method, invariance, velocity of light, barycenter, Solar system, semi-major axis, orbit

**Introduction**

Currently (February 2011), about 70% of candidates for extrasolar planets are defined on the basis of the method of radial velocities - spectroscopic measurements of radial velocity component of stars [Schneider].

The planet (or planets) is found out on a variation of radial velocity of parent star, observed from the Earth, - $\upsilon_r$, which in nonrelativistic case when $\upsilon_r$ is substantially less than the velocity $c$, detected radiation from the star, is determined by the Doppler formula

$$\upsilon_r(t) = -c\Delta f(t)/f_0. \tag{1}$$

Here $\Delta f(t) = f(t) - f_0$, $f(t)$ – registered frequency of radiation from a star of the selected spectral line corresponding with "laboratory" (at $\upsilon_r = 0$) frequency $f_0$. The value of the radial velocity is non-negative when the star is approaching the observer, and the approach of the star to the observer $\upsilon_r < 0$.

Registered with the time variation of the radial velocity of stars, in the case of neglecting the mutual gravitational perturbations of exoplanets interpolated superposition of Keplerian orbital elements of the $N$ hypothetical planets circulating about the star (strictly speaking, the center of mass of extrasolar system) [Ferraz-Mello, 2005]:

$$\upsilon_r(t) = \sum_{i=1}^{N} V_i(\cos[\omega_i + \theta_i(t, e_i, \lambda_i, P_i)] + e_i \cos\omega_i) + \sum_{k=1}^{l} C_k t^k. \tag{2}$$

The unknown quantities are determined by the criterion of minimum mean-square difference between the model radial velocity (2) and obtained from observations, are the eccentricities $e_i$, mean longitude $\lambda_i$, orbital periods $P_i$ of extrasolar planets, which can be expressed the true anomaly $\theta_i$, as well as the magnitude argument of pericenters of the planet's orbit $\omega_i$, the coefficients $C_k$ of polynomial trend of star's radial velocity due to the limited duration of observation, the amplitude coefficients $V_i$, associated with minimal masses of exoplanets. Received currently, using radial velocity based on the Doppler Effect, the orbital parameters and masses of extrasolar planets lead to paradoxical orbital configurations of precalculate exoplanet's orbits (Fig. 1). Moreover, so close orbital coexistence of stars and hot "Jupiters" is not realized on cosmogonic intervals of time.

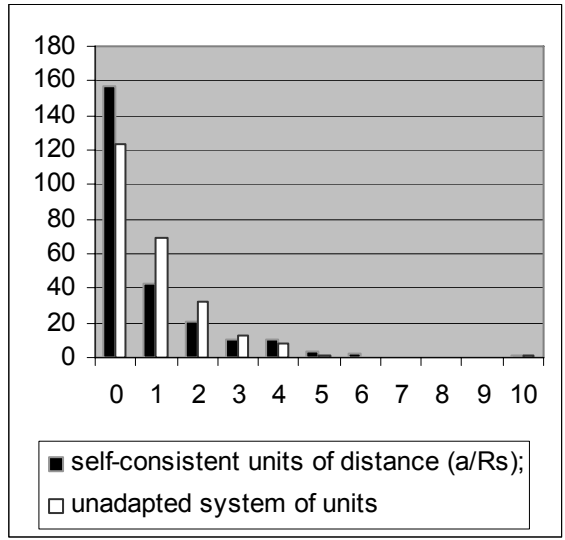

Fig. 1. Histogram of exoplanet orbit's semi-major axes: in absolute terms and in a self-consistent system of units (both systems are normalized to 10).

**The account of non-inertial movements**

As appears from a fundamental principle of the least action the condition of mechanical system is completely defined by the task of coordinates and velocities system component. However force interaction (the movement equation) in mechanical system are defined through derivative of velocities, that is accelerations (Lagrange's or Newton's equations do not contain derivatives of the accelerations) [Landau, 1973].

Doppler Effect, expressed in the form of (1), does not account for the force (gravitational) influence to center of mass of "a star – planets" dynamic system and characterized by the acceleration of the star, which spectrum is measured.

Without loss of generality, we will consider that variations of the radial velocity of star *S* are caused by its reference around the force center (center of mass *O* hypothetical extrasolar system) with a constant angular velocity $\Omega = \upsilon/r_0$ on a circular orbit with radius $r_0$ (fig. 2). We will consider that the observer is in a plane of an orbit of star *S* at a distance $r >> r_0$ from the center *O* about which the observer has a component of radial velocity $w_r(t)$. Assuming that $|w_r|$ and $\upsilon << c$, the clock in the system of coordinates associated with the star *S*, denoted by $t_S$, and in case of the observer – $t$ (at $\upsilon << c$ clocks are synchronized). Due to the significant distance *r* from the star to the observer the star *S* is a point source of radiation.

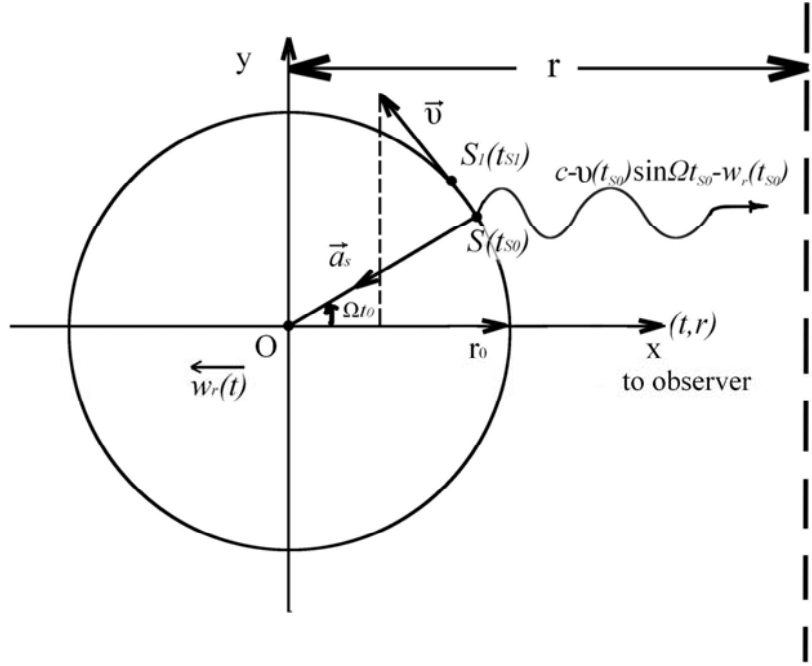

Fig. 2. Circular orbit of the star *S* with respect to the center of mass moving relative to the observer with the radial velocity $w_r(t)$.

Light from star *S*, generated at the time $t_{S0}$, taking into account the principle of Galilean relativity (the nonrelativistic case; justification below) to reach the observer at time

$$t_0 = \frac{r(t_{S0}) - r_0 \cos \Omega t_{S0}}{c - \upsilon(t_{S0}) \sin \Omega t_{S0} - w_r(t_{S0})}. \tag{3}$$

Radiation from a star, generated in $t_{S1}$ through the period of radiation $T_S$, reaches reach the terrestrial observer in $t_{II} = T_S + \dfrac{r(t_{S1}) - r_0 \cos \Omega t_{S1}}{c - \upsilon(t_{S1}) \sin \Omega t_{S1} - w_r(t_{S1})}$. (4)

Since $t_{S1}=t_{S0}+T_S$, $a_S=v^2/r_0$, that, entering designations $a_r(t_{S0})=a_S\cos(\Omega t_{S0})$, $\upsilon_r(t)=\upsilon(t)\sin\Omega t$, and considering that $\Omega T_s \ll 1$, $w_r=const$ during $T_s$, in the case under consideration $|\upsilon|=cost$, we have

$$\begin{aligned}\upsilon_r(t_{S1}) &= \upsilon_r(t_{S0})\cos(\Omega T_S) + \upsilon\sin(\Omega T_S)\cos(\Omega t_{S0}) \approx \upsilon_r(t_{S0}) + T_S a_r(t_{S0}),\\ r(t_{S1}) &= r(t_{S0}) + w_r T_S.\end{aligned} \tag{5}$$

Consequently, the period $T=t_{II}-t_0$ of radiation registered by the observer, periodicity $T_s$, of star *S* according to (3) - (5), taking into account that $r(t_{S0})=r \gg r_0$, with considered accuracy will be equal:

$$T = T_S\left(1 + \frac{\overline{w_r}}{c - \overline{w_r} - T_s a_r(t_{S0})} + \frac{a_S \cos(\Omega t_{S0}) r(t_{S0})}{(c-\overline{w_r})(c-\overline{w_r} - T_s a_r(t_{S0}))}\right), \tag{6}$$

where $\overline{w_r} = w_r + \upsilon_r(t_{S0})$.

And finally, given that in this case $\left|\overline{w_r}\right| \ll c$, $a_S T_S = \Omega^2 r_0 T_S \ll c$, we obtain:

$$\frac{T}{T_S} = 1 + \frac{\overline{w_r}}{c} + \frac{r}{c}\frac{a_r}{c}, \qquad (7)$$

so that in case of relevant frequencies, we have

$$\Delta f / f_S = -\overline{w_r}/c - r a_r / c^2, \qquad (8)$$

where $\Delta f = f(t) - f_S$.

When $a_r=0$ (case of uniform motion of the radiation (light) source) the expression (8) is equivalent to the well-known non-relativistic formula derived in 1842 by Doppler in relation to the phenomena of sound and are common for electromagnetic waves by Fizeau in 1848.

However at non-uniform movement of a source of radiation change of a registered frequency spectrum is caused also by presence radial component of acceleration of the emitting object (star), so that the change in frequency of the detected radiation from the star is at zero value of the radial velocity of a star. Thus the importance of the second composed in (8) just also is essential to distant sources, when $r/c \gg 1$, as it is realised, unlike Solar system, for hypothetical extrasolar planets.

Occurrence of the second composed in (8) is formal can be interpreted gravitational shift in the equivalent gravitational field of rapidly moving system.

Relation of the form (8), on the other hand, is a consequence of the developed back in 1908 by Ritz Emission (ballistic) theory, in particular, the electrodynamics processes [Ritz, 1911]. According to this theory, based on the principle of Galilean relativity, the light is a stream of specific particles (reons) emitted by the velocity of light.

Erroneous estimations about absence of evidence of Ritz emission theory in binary star systems, given the Dutch astronomer De Sitter in 1913 and later [De Sitter, 1913.1924], and repeating them in various monographs and textbooks [Landsberg 1976; Sivukhin, 1980] did not contribute to the development of emission theory and its practical application. However, actually, arguments resulted till now as objections against Ritz theory are not justified, but by deeper consideration actually are acknowledgement of the given theory [Fox, 1965, Eljashevich, 1995; Moon, 1953, Martinez, 2004]. This theory is consistent with the electron theory of dispersion, explains the observed perihelion of Mercury, the effects of microlensing, the life expectancy π-and μ-mesons, quantitatively satisfy the transverse Doppler Effect, the phenomenon of aberration of galaxies, interferometry observations of binary stars. Under this theory holds the principle of local irreversibility remains the law of conservation of energy and momentum for a period of time between emission and absorption of light by a substance, as well as an opportunity to resolve the so-called problem of dark matter and a number of other hitherto unsolved problems.

Let's estimate amplitude composed in the expression (8), caused by radial acceleration of a star at which the spectrometer method carries out search extrasolar planets.

Let's assume that the star with mass $M_S$ have one planet with a mass $M_p$, comparable with the relative mass of Jupiter in our Solar system, so that $M_p/M_S=10^{-3}$ ($M_p=1.9\times10^{27}$ kg). Taking into account characteristic distances to stars, which are currently detected extrasolar planets candidates, we accept distance to an estimated star of equal 150 light years ($\approx$46 pc), so $r/c \approx 4.74 \times 10^9$ sec (s).

The distance from the center of mass of the star to the forcer center $O$ (the center of mass of the star-planet system), within the limits of our model in Figure 2, is determined by the obvious relation:

$$r_0 = \frac{M_p/M_S}{1+M_p/M_S}\rho_{PS}, \qquad (11)$$

in which $\rho_{pS}$ – the distance between the centers of mass of the planet and star.

By solving the two bodies problem follows that movement of bodies and, in particular, star $S$ in the barycentric coordinate system with reference point at the center of mass ($O$) system $P$-$S$ and with invariable directions of the coordinate axes (in general $OXYZ$) is described by the equations and, consequently, solutions similar to the case of the relative motion of the bodies $P$ and $S$, but with formal replacement of factor $\mu=G(M_S+M_P)$ to [Duboshin, 1975]

$$\mu = \frac{GM_p^3}{(M_S+M_p)^2}, \qquad (12)$$

where $G=6.67\times 10^{-11}$ m$^3$/(kg$\times$s$^2$) – the gravitational constant. Therefore, the frequency $\Omega$ circulation in a circular orbit of radius $a=r_0$ with respect to the barycenter $O$ (equivalent to the force center with reduced mass $\mu/G$) will be equal

$$\Omega = \sqrt{\mu} r_0^{-3/2} \qquad (13)$$

and, consequently, to a maximum value of radial acceleration, taking into account (11) - (13), we obtain the obvious expression: $a_S = \Omega^2 r_0 = GM_p/\rho_{PS}^2$. (14)

According to (14) for the planet, farthest from the star at a distance $\rho_{pS}$=10 a.u., the maximum velocity component $v_a=(a_r)r/c$ in the second term (8), due to the acceleration of the star is equal to 268.4 m/s, and in the case $\rho_{pS}$=1 a.u. we have $v_a=2.68\times 10^4$ m/s. The amplitude of the velocity component $v_a$ depending on the distance of the planet to the star for different values of the mass of a hypothetical extrasolar planet are shown in Figure 3.

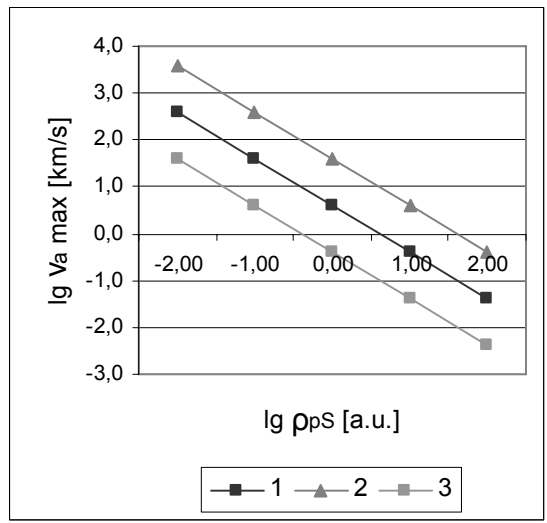

Fig. 3. The amplitude of the velocity component $v_a$ depending on the distance $\rho_{pS}$ to star, whose distance from the observer assumed to be 150 light years. In the figure:

1 - planetary mass $M_P$ is the mass of Jupiter, $M_J$, 2 – $M_P$ equal to 10 $M_J$ 3 - $M_P$ is 0.1 $M_J$.

With increasing distance from the observer to the star, as follows from (8), the amplitude of the velocity components $v_a$ proportional increases. Thus its size depends on absolute value of mass of a planet and inversely proportional to a square of distance from a planet to a star.

Hence (at domination of the second composed in (8)), for equidistant from the observer stars should not be any registered selection on mass of stars at the same shift of spectral frequencies of radiation of stars. While the selective effect depending on remoteness of a planet from a star (the closer the planet to a star, is easier to find out it) should be shown. Thus, it appears, the further from the observer there is a star, the it is easier to find out, with other things being equal, a spectroscopic method in it a planet (planets)!

Currently available statistical observational data, taking into account the renormalization of the array of samples studied stars, confirms this pattern (Fig. 4).

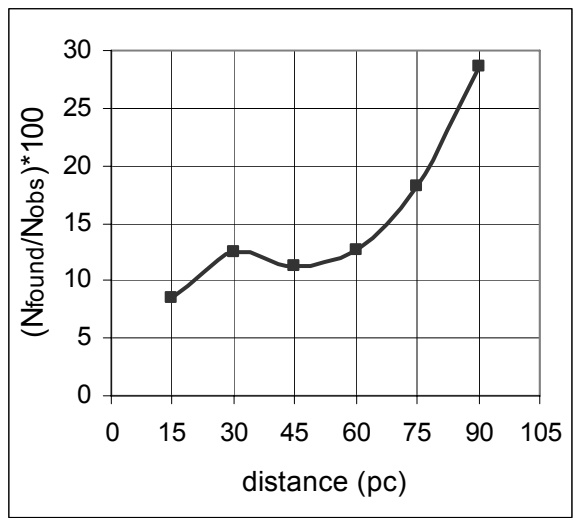

Fig. 4. With increasing distance $r$ to study the probability of stars (relative to the investigated sample of stars, the number of the found planets $N$) detected candidate extrasolar planets increases.

Considering that found in the present spectroscopic method without the second term in (8) extrasolar planet candidates correspond to deviations of radial velocity component $w_r$ from tens km/s to hundreds m/s, it follows from the results shown in Figure 3, in this case in this study real boundary of the existence of extrasolar planets is removed on the degree, it any more units, but tens of astronomical units, which, in turn, is consistent with the modern theory of formation and the dynamical evolution of planetary systems.

The resulted estimations essentially will not change at an elliptic orbit of star $S$ about the barycenter $O$. In this case, the acceleration $a_S$, still, will be directed to barycenter $O$, which will be one of the focuses of the elliptical orbit, and time-varying velocity vector $v_S$ will not be perpendicular to the radius vector $\rho_S$ (Figure 5).

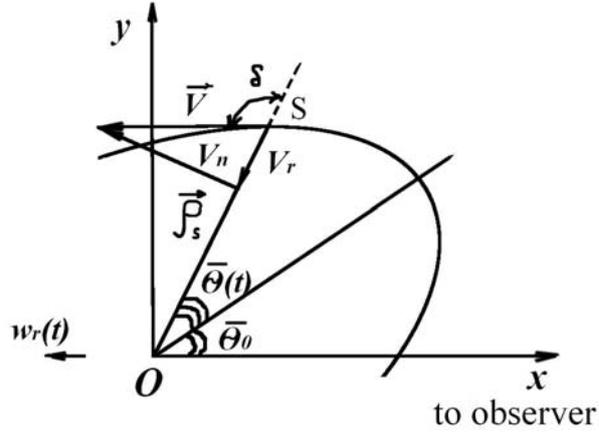

Fig. 5. The elliptical orbit of the star S with respect to the barycenter $O$ of the star-planet.

In parities (3) - (8) and (14) in case of an elliptic orbit it is necessary to carry out only formal replacements $r_0$ to $\rho_S$ and $\upsilon_r(t) = V_\rho \cos(\theta + \theta_0) + V_n \sin(\theta + \theta_0)$, angle $\theta_0$ defines position of a line of apses concerning a direction to the observer.

Note that the effects caused by the second term in (8) for objects in the solar system are negligible. For observed in area $r$=100 a.u. Kuiper belt objects $a_K = GM_C/\rho_{CK}^2 \approx 5.9\times10^{-7}$ m/s², consequently, the amplitude of the corresponding component $\upsilon_{aK}$ in (8) we obtain a value $\upsilon_{aK}$=2.9 sm/s, , which on five degree less than average orbital speed of these objects in Solar gravitational field.

**Some consequences and estimations**

Considered above the amendment to registered frequencies of star's spectrum for spectrometric measurements in exoplanet search programs are caused along with the account non-inertial movements of a radiation source (star) and the assumption of a constancy a velocity of light concerning a radiation source, unlike a postulate in the special theory of a relativity (SRT) about a constancy of light velocity concerning the observer.

Let's show further that the applied assumption practically is not essential at optical (photometric) monitoring of the studied stars, that have exoplanets (planetary systems).

As can be seen from figure 2, the timeinterval that light reaches the observer is given by (3). Take the clock in the reference system associated with radiation from the star is equal to $t_S$, and in the reference frame of observer – $t_н$, taking into account that $\upsilon/c$, $w_r/c$ and $r_0/r \ll 1$ (clock are synchronized), to the clock of observer obtain: $t_н = t_S + t_0$. (15)

Implementation of the transformation, similar to relations (4) - (7), and let $\varphi=\Omega(t_н-r/c)$ the value of the angular position of the star in the barycentric orbit, corrected, taking into account the finite velocity of light, we have $\varphi = \arcsin(y/r_0) + \Phi_0(y/r_0)$, (16)
where $y/r_0=\sin(\Omega t_S)$, $\Omega=2\pi/\tau$, $\Phi_0=\Omega(r/c)(\upsilon/c)$, $\tau$ – stellar orbital period (around barycenter of star-planet system).

From (16) implies that the extreme value of function $\varphi(y/r_0)$ is reached at $\Phi_0=[1-(y/a)^2]^{-1/2}$, and since $\Phi_{0\text{к}}=1$, at the further increase $\Phi_0$ for the observer there will be multiple images of a star - three stars in various points on the same barycentric orbit.

In the case of values $\Phi_0$ less than one multiple images of the stars should not arise. If $\Phi_0 \ll 1$, then according to (16) barycentric motion of the star seen by the observer, will be slightly different from sinusoidal.

Given the previously cited estimates for representing the current interest configurations of barycentric orbits of stars in a search program for exoplanets, at distances up to $10^3$ light years it is had the maximum estimation (at $\rho_{PS}=0.1$ a.u.) $\Phi_0=6.6\times10^{-2} < \Phi_{0\text{к}}$, so any multiple images of an investigated star at the expense of distinctions in a velocity of light on various sites of the barycentric it should not to be observed.

It is similar, easy to show that for small values of $\Phi_0$ changes visible brightness and the apparent orbit of the stars studied in the existing programs of the search for exoplanets are unimportant [Moon, 1953]. It should also be noted that for visual and spectroscopic binary stars, the Cepheids, for the assumption of constancy of the velocity of light relative to the radiation source does not lead to any fundamental mismatch with the observations, i.e. does not allow us to choose between alternatives: the constancy of the velocity of light relative to the radiation source, or - an observer [Moon, 1953 Eneev, 1981 a, b; Belopol'skii, 1954].

If the observer does not settle down in a plane of barycentric orbits of the star, then as given in (8) components of the radial velocity and acceleration should take appropriate projection on the plane perpendicular to the "picture plane" (normal is passed through the barycentre of the planetary system and the observer) $\upsilon_r(i)=\upsilon\sin(\Omega t)\cos(i)$, $a_r(i)=a_S\cos(\Omega t)\cos(i)$, where $i$- the angle between the observer (perpendicular to the plane of the sky) and the plane of barycentric orbits the star. The ratio between the amplitudes of the two components in (8) at $w_r=0$ does not change, but the components themselves are not identically zero, except when the barycentric orbit of the star is located in the plane ($i=90^0$).

Statistics (February 2011) on the detected candidate extrasolar planets in the system indicate that the total number of extrasolar planets based on the results of spectral measurements is 368. Including 46 planets (12.5%) also confirmed astrometric method, and 26 exoplanets (7%) confirmed "transit method" [Schneider], i.e. the total number of candidates in exoplanets that were duplicated in two ways (one of them is spectrometry) is 72 planets (19.6%). In this case, of the 72 extrasolar planets in 12 (16.7%), for which data are published, the parameters of exoplanets (mass planets and stars, the orbital parameters) are improved. In addition, the total number of candidates for extrasolar planets detected by RV method were also clarified options 30 exoplanets (8.1%), and of the 110 extrasolar planets for which data are obtained by the transit method is subsequently refined parameters 6 planets (5.5%).

Comparison of the data of the incorporated exoplanet catalogue with earlier given (2002, 2006) has revealed a divergence in ~ 5 % of cases. Values of the orbital periods of four exoplanets in the compared catalogues differ in several times and at one of planets difference of the orbital period reaches 3.5 times.

The total not included for any reasons in the incorporated catalogue of planets has made 134 candidates (20.3 %), from them in 8 samples of a variation of radial velocity have been caused by a star origin, more than 50 candidates have been recognized by needing reception of the new data, and for one exoplanet different groups of observers have received essentially different estimations.

Thus, the data available now (19.6 %) on comparison of results of a spectrometric method with alternative methods (transit or astrometry) are not still statistically significant. However in process of increase in a file of the compared data, under condition of application correct (a priori not coordinated) independent methods (in relation to a spectral method) detection, the results resulted in the present article will serve as the reliable test for check of invariance of alight velocity.

### Conclusion

Formally, the manifestation in the recorded frequency spectrum components caused by the presence of radial acceleration of the star relative to the barycenter of the stellar system can be interpreted in the framework of general relativity with an appropriate representation of the gravitational potential for the case of accelerated dynamic system. At the same time, the presence in the recorded frequency spectrum of the input beam components of the acceleration of the emitting object (a star), even at a zero value of radial velocity of stars, the most consistently substantiated in the framework of the emission (ballistic) theory by Ritz.

Since the component of the barycentric radial acceleration is proportional to the distance to the emitting object (a star), its contribution, in contrast to the solar system, is essential for extrasolar planets.

For a star which are from the observer on distance $r$ and moving on circular barycentric orbit with frequency $\Omega$ (in the case of a single or dominant mass of the planet equal to frequency of the reference extrasolar planet), the spectral shift due to the accelerated movement surpasses corresponding classical Doppler Effect in $\Omega r/c$ times, so if the quantity of light years to an investigated star exceeds in $2\pi$ times reduced exoplanet's orbital pediod, expressed in years, the radial component of acceleration is dominant, which is realized for all candidates of extrasolar planets discovered till now by a spectrometer method.

Due to the dependence of radial component of acceleration from the absolute values of the mass of the planet for equidistant from the observer, the stars should not show the effect of selection on the masses of stars at the same bias detected spectral frequencies of the stars, because relatively close to the Solar system, stars can ignore the influence of the galactic component of radial acceleration due to treatment of the stars relative to the center of mass of the Galaxy.

The further from the observer the star, the, with other things being equal, settles down the spectrometer method appears to find out easier in it the planets, therefore the corresponding trend is observed in correct histograms.

Presence of essential effects in case of display in a registered frequency spectrum components of radial acceleration of a star rather barycenter of star-planet system in process of occurrence of statistically significant comparisons of the results of the spectral method with alternative methods of detecting extrasolar planets will get a reliable test for checking the invariance of the light velocity.

**References**

1. Schneider J. (CNRS-LUTH, Paris Observatory) // http://exoplanet.eu/catalog.php
2. Ferraz-Mello S., Michtchenko T. A., Beauge C., Callegari N. Extra-solar planetary systems // Lect. Not. Phys. 2005. V. 683. P.219-271.
3. Bonfils X., Mayor M. et al. The HARPS search for southern extra-solar planets // Astr. and Astroph. 2007. V. 474. P. 293-299.
4. Scargle J. Studies in astronomical time series analysis statistical aspects of spectral analysis of unevenly spaced data // Ap. J. 1982. V.263. P. 835-853.
5. Schwarzenberg-Czerny A. Fast and statistically optimal period search in uneven sampled observations // Ap. J. 1996. V. 460. P. L107-L110.
6. O'Toole S., Tinney C., Jones H. The impact of stellar oscillations on Doppler velocity planet searches // MNRAS. 2008. V. 386. P. 516-520.
7. Wright J. Radial velocity jitter in stars from the California and Carnegia planet search at Keck observatory / / PASP. 2005. V.117. P.657-664.
8. Ida S., Lin D. Towards a deterministic model of planetary formation // Astroph. J. V. 604. N1.2004. P. 388-413.
9. Safronov V.S., Vityazev A.V. Origin of the solar system: Progress in science and technology. Ser. Astronomy. T.24. M.1983;
10. Mushailov B.R., Ivanovskaya L.M., Teplitskaya V.S. Homogeneous statistical distributions of extrasolar planets by their dynamic parameters, Cosm. Research. 2010. T.48. № 4. P. 380-384.
11. Landau L.D. and Lifshitz E.M. Mechanics. Ser. Theoretical Physics. V.1. M. 1973. 208 pp.
12. Frankfurt U.I., Frank A.M. Optics of moving bodies. M. Science .1972. P.136.
13. Bömmel H. Measurement of the frequency shift of gamma rays in accelerated systems using the Mässbauer effect. "Mossbauer effect". Ld. 1962. P. 229-232.
14. Pound, R. On the weight of photons // Physics-Uspekhi. T. 72. № 12. 1960. P.673-683.
15. Ritz W. Gesammelte werke. Walter Ritz. Oeuvres. Paris: Gautier-Villars. 1911. P. 317-492.
16. De Sitter W. Ein astronomischer beweis für die konstanz der lichtgesshwindigkeit // Proc. Acad. Amster. 1913. N 15. PP 1297-1312; Phys. Z. 1913. N 14. P. 429.
17. De Sitter W. Über die Genauigkeit, innerhalb welcher die Unabhängigkeit der Lichtgeschwindigkeit von der Bewegung der Quelle behauptet werden kann. // Physik. Zeitschr. XIV, 1913. P. 1267; Bull. Astron. Inst. Netherlands. N2. 1924. P. 121, 163.
18. Landsberg G.S. Opt. M. 1976. P. 451.
19. Sivukhin D.V., General Course of Physics. T. IV. Opt. M. 1980. 629 p.
20. Fox J. Evidence against emission theories // American J. of Physics. V. 33. N1. 1965. PP. 1-17.
21. El'yashevich M.A., Kembrovskaya N.G., Tomilchik L.M. Walter Ritz and his research on the theory of atomic spectra, Physics-Uspekhi. № 4. 1995. P. 457-480.
22. Moon P., Spencer D. Binary stars and the velocity of light // J. of the optical society of America. V. 43. N.8. 1953. P 635-641.

23. Martinez A. Ritz, Einstein and emission hypothesis // Physic in perspective. N.6. 2004. P. 4-28.
24. Duboshin G.N. Celestial mechanics. Main objectives and methods. M. Science. 1975. 800.
25. Sebehey, Theory of orbits. M. 1982. 656.
26. Gerasimov, I.A., Vinnikov E.L., Mushailov B.R. Canonical equations of celestial mechanics. M. 1996. 208 pp.
27. Landscheidt T. Beziehungen Zwischen der sonnenaktivität und dem massenzentrum des Sonnensystems. Nachr. D. Olbersgesellschaft Bremen. N.100. 1976. P.3-19; Klimavorhersage mit astronomischen Mitteln. Aus Fusion. N.1. 1997.
28. Standish E.M.; Williams J.G. (1992), "Orbital Ephemerides of the Sun, Moon, and Planets", in Seidelmann, P. Kenneth, Explanatory Supplement to the Astronomical Almanac (1 ed.), Mill Valley, CA: University Science Books, pp. 279-323, ISBN 0-935702-68-7, http://books.google.com/books?id=uJ4JhGJANb4C&pg=PA279.
29. Eneev T.M., Kozlov N.N. model of the accumulation process of forming planetary systems. I. Numerical experiments, Astron. Herald. 1981. T. 15. № 2. P.80-94. II. The rotation of the planets and the connection with the theory of gravitational instability, Astron. Bulletin 1981. T. 15. № 3. P. 131-141.
30. Belopol'skii A.A. Astronomical works. (Classics of science. Mathematics. Mechanics. Physics. Astronomy.). M: Gostekhteoretizdat, 1954.
31. Semikov S., SRT hundred years. Is there an alternative? // Engineer. 2005. N 11. S. 21-24.
32. http://www.phys.unsw.edu.au/ ~ cgt / planet / Targets.html
33. http://feps.as.arizona.edu/observations.html